\pdfoutput=1
\PassOptionsToPackage{hidelinks, bookmarks=false}{hyperref}

\makeatletter
\@namedef{ver@hyperxmp.sty}{9999/12/31}
\makeatother

\documentclass[sigconf, natbib=true, anonymous=false]{acmart} 

\AtBeginDocument{%
  }
\usepackage[utf8]{inputenc} 
\usepackage[T1]{fontenc}    
\usepackage{url}            
\usepackage{nicefrac}       
\usepackage{microtype}      
\usepackage{bm}
\usepackage{bbm}
\usepackage{todonotes}
\usepackage{algorithm}
\usepackage{algpseudocode}
\usepackage{multicol}
\usepackage{caption}
\usepackage{multirow}
\usepackage{makecell}
\usepackage{booktabs}           
\usepackage{amsfonts}           
\usepackage{graphicx}           
\usepackage{bold-extra}
\usepackage{amsmath, mathtools, bm}
\usepackage{colortbl}
\definecolor{Gray}{gray}{0.9}

\begin{document}

\title{CRAB: Codebook Rebalancing for Bias Mitigation in Generative Recommendation}

\author{Ziheng Chen}
\authornote{Both authors contributed equally to this research.}
\email{albertchen1993pokemon@gmail.com}
\affiliation{%
  \institution{Walmart Global Tech}
  \city{Sunnyvale}
  \country{USA}
}

\author{Zezhong Fan}
\authornotemark[1] 
\email{zezhong.fan@walmart.com}
\affiliation{%
  \institution{Walmart Global Tech}
  \city{Sunnyvale}
  \country{USA}
}

\author{Luyi Ma}
\email{luyi.ma@walmart.com}
\affiliation{%
  \institution{Walmart Global Tech}
  \city{Sunnyvale}
  \country{USA}
}

\author{Jin Huang}
\email{jin.huang@stonybrook.edu}
\affiliation{%
  \institution{Stony Brook University, NY, USA}
  \city{Stony Brook}
  \country{USA}
}

\author{Lalitesh Morishetti}
\email{lalitesh.morishetti@walmart.com}
\affiliation{%
  \institution{Walmart Global Tech}
  \city{Sunnyvale}
  \country{USA}
}

\author{Kaushiki Nag}
\email{kaushiki.nag@walmart.com}
\affiliation{%
  \institution{Walmart Global Tech}
  \city{Sunnyvale}
  \country{USA}
}

\author{Sushant Kumar}
\email{sushant.kumar@walmart.com}
\affiliation{%
  \institution{Walmart Global Tech}
  \city{Sunnyvale}
  \country{USA}
}

\author{Kannan Achan}
\email{kannan.achan@walmart.com}
\affiliation{%
  \institution{Walmart Global Tech}
  \city{Sunnyvale}
  \country{USA}
}

\newcommand{\X}{\mathcal{X}}
\newcommand{\Y}{\mathcal{Y}}
\newcommand{\R}{\mathbb{R}}
\newcommand{\paramspace}{\boldsymbol{\Theta}}
\newcommand{\params}{\boldsymbol{\theta}}
\newcommand{\model}{f_{\params}}
\newcommand{\modeledit}{f_{\params'}}
\newcommand{\inst}{\boldsymbol{x}}
\newcommand{\cfinst}{\boldsymbol{\widetilde{x}}}
\newcommand{\pred}{\hat{y}}
\newcommand{\instedit}{\inst_c}
\newcommand{\instq}{\inst_q}
\newcommand{\labeledit}{y_c}
\newcommand{\labelq}{y_q}
\newcommand{\prededit}{\hat{y}_c}
\newcommand{\editorparamspace}{\boldsymbol{\Psi}}
\newcommand{\editorparams}{\boldsymbol{\psi}}
\newcommand{\dataset}{\mathcal{D}}
\newcommand{\testset}{\dataset_{q}}
\newcommand{\datasetedit}{\dataset_c}
\newcommand{\loss}{\ell}
\newcommand{\Loss}{\mathcal{L}}
\newcommand{\method}{\textsc{FROG}\xspace}
\newcommand{\softmax}{\text{softmax}}

\newcommand{\Prob}{\mathbb{P}}
\newcommand{\Z}{\mathbb{Z}}
\newcommand{\E}{\mathbb{E}}
\newcommand{\G}{\mathcal{G}}
\newcommand{\advG}{\widetilde{\G}} 
\newcommand{\Gobs}{\G^{\text{obs}}}
\newcommand{\U}{\mathcal{U}}
\newcommand{\I}{\mathcal{I}}
\newcommand{\V}{\mathcal{V}}
\newcommand{\Vnew}{\V^{\text{new}}}
\newcommand{\Vadv}{\V^{\text{adv}}}
\newcommand{\edges}{\mathcal{E}}
\newcommand{\edgesnew}{\edges^{\text{new}}}
\newcommand{\edgesobs}{\edges^{\text{obs}}}
\newcommand{\edgesadv}{\edges^{\text{adv}}}
\newcommand{\bgraph}{\G=(\U, \I, \edges)}
\newcommand{\neigh}{\mathcal{N}}
\newcommand{\adjM}{A}
\newcommand{\advadjM}{\widetilde{A}}
\newcommand{\adjMij}{{A}_{i,j}}
\newcommand{\train}{\dataset_{\text{train}}}
\newcommand{\test}{\dataset_{\text{test}}}
\newcommand{\features}{\mathcal{X}}
\newcommand{\labels}{\mathcal{Y}}
\newcommand{\hypspace}{\mathcal{H}}
\newcommand{\w}{\bm{\omega}}
\newcommand{\h}{\bm{h}}
\newcommand{\advh}{\widetilde{\bm{h}}}
\newcommand{\hyp}{h_{\params}}
\newcommand{\gnn}{g(\adjM, \X; \W)}
\newcommand{\Cons}{\mathcal{L}_{\text{cstr}}}
\newcommand{\ladv}{\ell_{\text{adv}}}
\newcommand{\ldist}{\ell_{\text{dist}}}
\newcommand{\lnew}{\ell_{\text{new}}}
\newcommand{\LF}{\Loss_{fa}}
\newcommand{\LC}{\Loss_{cf}}
\newcommand{\ind}{\mathbbm{1}}
\newcommand{\inputs}{\mathcal{V}}
\newcommand{\outputs}{\mathcal{Y}}
\newcommand{\insta}{X_i}
\newcommand{\weights}{\boldsymbol{\omega}}
\newcommand{\forget}{\mathcal{D}_f}
\newcommand{\remain}{\mathcal{D}_r}
\newcommand{\all}{\mathcal{D}}
\newcommand{\LLMori}{\mathcal{M}_{\theta}}
\newcommand{\un}{\mathcal{M}_{un}}
\newcommand{\optimal}{\mathcal{\mathcal{M}}_{\theta^{*}}}
\newcommand{\indep}{\perp \!\!\! \perp}
\newcommand{\cnf}{\bold{Z_{cnf}}}
\newcommand{\graph}{\mathcal{G}}
\newcommand{\node}{v}
\newcommand{\edge}{\mathcal{E}}
\newcommand{\nshare}{\boldsymbol{\Tilde{\theta}_{sh}}}
\newcommand{\nforget}{\boldsymbol{\Tilde{\theta}_{f}}}
\newcommand{\nretain}{\boldsymbol{\Tilde{\theta}_{r}}}
\newcommand{\selshare}{\boldsymbol{\Hat{\theta}_{sh}}}
\newcommand{\selforget}{\boldsymbol{\Hat{\theta}_{f}}}
\newcommand{\selretain}{\boldsymbol{\Hat{\theta}_{r}}}
\newcommand{\orishare}{\boldsymbol{\theta}_{sh}}
\newcommand{\oriforget}{\boldsymbol{\theta}_{f}}
\newcommand{\oriretain}{\boldsymbol{\theta}_{r}}
\newcommand{\gradfshare}{\mathbf{g}_{sh}^{F}}
\newcommand{\gradfforget}{\mathbf{g}_{f}^{F}}
\newcommand{\gradfretain}{\mathbf{g}_{r}^{F}} 
\newcommand{\gradrshare}{\mathbf{g}_{sh}^{R}}
\newcommand{\gradrforget}{\mathbf{g}_{f}^{R}} 
\newcommand{\gradrretain}{\mathbf{g}_{r}^{R}}
\newcommand{\wforget}{\omega_f}
\newcommand{\wretain}{\omega_r}
\newcommand{\inpori}{\mathbf{x}_{u}}
\newcommand{\inpcf}{\mathbf{x}_{u}^{*}}

\newcommand{\Uset}{\mathcal{U}}
\newcommand{\Iset}{\mathcal{I}}
\newcommand{\LLM}{\mathcal{M}}


\newcommand{\xg}[1]{\todo[inline,color=red!60]{\textbf{xiangguo:} #1}}
\newcommand{\xgr}[1]{\textcolor{red}{\textbf{xiangguo:} #1}}

\newcommand{\zezhong}[1]{\todo[inline,color=pink!60]{\textbf{zezhong:} #1}}
\newcommand{\luyi}[1]{\todo[inline,color=green!60]{\textbf{luyi:} #1}}
\newcommand{\jiali}[1]{\todo[inline,color=orange!60]{\textbf{Jiali:} #1}}
\newcommand{\ziheng}[1]{\todo[inline,color=blue!60]{\textbf{Ziheng:} #1}}
\newcommand{\hadi}[1]{\todo[inline,color=yellow!60]{\textbf{Hadi:} #1}}
\newcommand{\sun}[1]{\todo[inline,color=cyan!60]{\textbf{sun:} #1}}
\newcommand{\yunzhi}[1]{\todo[inline,color=magenta!60]{{\bf Yunzhi:} #1}}

\begin{abstract}
Generative recommendation (GeneRec) has introduced a new paradigm that represents items as discrete semantic tokens and predicts items in a generative manner. Despite its strong performance across multiple recommendation tasks, existing GeneRec approaches still suffer from severe popularity bias
and may even exacerbate it. In this work, we conduct a comprehensive empirical analysis to uncover the root causes of this phenomenon, yielding two core insights: 1) imbalanced tokenization inherits and can further amplify popularity bias from historical item interactions; 2) current training procedures disproportionately favor popular tokens while neglecting semantic relationships among tokens, thereby intensifying popularity bias.
\par\indent
Building on these insights, we propose CRAB, a post-hoc debiasing strategy for GeneRec that reduces popularity bias by rebalancing semantic token frequencies. Specifically, given a well-trained model, we first identify and split over-popular tokens while preserving the overall hierarchical structure of the codebook. Based on the adjusted codebook, we further introduce a hierarchical regularizer to enhance semantic consistency, encouraging more informative representations for unpopular tokens during training. Experiments on real-world datasets demonstrate that CRAB effectively alleviates popularity bias while maintaining competitive performance.
\end{abstract}
\settopmatter{printacmref=false} 
\setcopyright{none}               
\renewcommand\acmConference[4]{}  
\maketitle

\section{Introduction}
\label{sec:intro} 
Generative recommendation (GeneRec) ~\cite{wang2024learnable, han2025mtgr, liu2025onerec} has emerged as a promising paradigm for sequential recommendation, demonstrating strong empirical performance across diverse domains~\cite{deng2025onerec,kong2025minionerec,rajput2023recommender}. It formulates recommendation as a sequence-to-sequence generation task: a tokenizer maps each item into a sequence of discrete codebook tokens ~\cite{yu2021vector}, and the model autoregressively predicts the next item based on the concatenated tokenized interaction history ~\cite{lv2024semantic}. This design enables modeling over a compact token vocabulary, facilitating efficiency and adaptation to new items. Moreover, powerful sequence models such as LLMs can be naturally incorporated to enhance long-sequence modeling performance.

Despite their strong performance, GeneRec often exhibit popularity bias. To quantify this effect, we follow~\cite{lu2025dual} and define the top $20\%$ of items by historical interaction frequency as popular items, which collectively account for a disproportionately large share of interactions, and the remaining items as unpopular ones. As shown in the left part of Figure~\ref{fig:Motivation}, generative recommenders tend to recommend popular items more frequently than traditional methods. Specifically, compared with SASRec, the recommendation frequency of popular items increases by $6.7\%$ and $7.2\%$ for Mini-OneRec (MOR) ~\cite{kong2025minionerec} and TIGER ~\cite{rajput2023recommender}, respectively, while the frequency for unpopular items decreases by $3.8\%$ and $4.1\%$. 

However, existing debiasing methods for LLM-based recommendation~\cite{lu2025dual,jiang2024item} do not generalize well to GeneRec. The key reason is that these approaches focus solely on mitigating bias at the model level, while overlooking the bias inherited and amplified by over-popular tokens in the codebook (see Section~\ref{sec:motivation}). Although some methods attempt to construct balanced codebooks~\cite{kuai2024breaking,deng2025onerec,hui2025semantics}, they typically constrain the number of items assigned to each token to prevent highly concentrated mappings, rather than addressing token popularity driven by historical interactions.

In this paper, we propose CRAB, a post-hoc approach for mitigating popularity bias in GeneRec. CRAB decouples the debiasing process into two stages: \textit{Codebook Rebalancing} and \textit{Hierarchical Semantic Alignment}. In the first stage, we identify over-popular tokens in the existing codebook and split them while preserving the hierarchical semantic structure of the remaining tokens; this process is formulated as a regularized K-means problem. In the second stage, we introduce a hierarchy-aware regularizer to enhance the representation of unpopular tokens by leveraging richer supervision signals from the semantics induced by the codebook. Overall, our key contributions are summarized as follows:

\begin{itemize}
\item \textbf{Problem:} We present the first investigation demonstrating that imbalanced codebooks give rise to over-popular tokens, which in turn bias the model toward popular items containing these tokens, thereby exacerbating popularity bias.


\item \textbf{Method:} We propose CRAB, an effective and efficient framework for mitigating popularity bias in GeneRec. It identifies and splits over-popular tokens in the codebook, and further enhances the representations of unpopular tokens through a hierarchical semantic regularizer.
\item \textbf{Performance:} Extensive experiments on real-world datasets demonstrate that CRAB reduces popularity bias by \textbf{16.5\%} while improving recommendation performance.
\end{itemize}

\section{Background and Preliminaries}
\label{sec:Preliminary}
We consider the sequential recommendation task.
Let $\mathcal{U}$ denote the set of users and $\mathcal{I}$ the set of items. Given a user $u \in \mathcal{U}$ with a chronologically ordered interaction history $\mathcal{H}_u = \{i_1, \ldots, i_T\}$, the goal is to predict the next item $i_{T+1}$.
Following recent GeneRec frameworks, each item is first converted into a sequence of discrete semantic tokens from its textual description via residual quantization methods (RQ-KMeans or RQ-VAE). An LLM is then trained to generate the token sequence of the next item.
\subsection{Residual Quantization}
For each item $i \in \mathcal{I}$, its textual description is first passed through a frozen encoder to obtain a continuous embedding $\boldsymbol{z}_i$. We aim to quantize it using $L$ hierarchical tokens with $L$ codebooks in a coarse-to-fine generation manner. At the $l$-th level, the codebook $\mathcal{C}^{l}=\left\{\boldsymbol{c}_{k}^{l} | k=1,\cdots K\right\}$ consists of $K$ codeword embeddings $\boldsymbol{c}_{k}^{l}$. When $l=1$, the initial residual is defined as $\boldsymbol{r}_{i}^{1}=\boldsymbol{z}_i$. Then at $l$-th level, the item token $s_i^{l}$ can be defined as the index of the closest codeword embedding from the codebook $\mathcal{C}^{l}$, and the residual at $l+1$-th level $\boldsymbol{r}_{i}^{l+1}$ is updated as follows:
\begin{equation}
\label{eq:RQ}
\begin{aligned}
s_{i}^{l}=\arg\min_{k}||\boldsymbol{r}_i^{l}-c^{l}_{k}||^{2},\quad \boldsymbol{r}_{i}^{l+1}=\boldsymbol{r}_{i}^{l}-\boldsymbol{c}^{l}_{s_{i}^{l}}
\end{aligned}
\end{equation}
The above process is repeated recursively $L$ times to get a tuple of $L$ codewords that represent the Semantic ID(SID) for item $i$. In RQ-KMeans, the codeword embedding $\boldsymbol{c}_k^{\,l} \in \mathcal{C}^l$ is defined as the centroid of the $k$-th cluster obtained by applying K-means to the residual set $\mathcal{R}^l = \{ \boldsymbol{r}_i^{\,l} \mid i \in \mathcal{I} \}$ at level $l$~\cite{wei2025cofirec}. 
\subsection{Autoregressive Generation}
By applying the residual quantization to each item in $\mathcal{H}_u$, the input is transformered into a flattened token sequence $X$. Accordingly, the target next item $i_{T+1}$ is also encoded as $Y$: 
\[\boldsymbol{X} = \big[
\underbrace{s_1^{1}, s_1^{2}, \ldots, s_1^{L}}_{i_1},
\ldots,
\underbrace{s_T^{1}, s_T^{2}, \ldots, s_T^{L}}_{i_T}
\big], \quad\boldsymbol{Y}=[\underbrace{s_{T+1}^{1},s^{2}_{T+1},\cdots s^{L}_{T+1}}_{i_{T+1}}]
\] where $s^{l}_i$ denotes the $l$-th token of item $i$. The LLM is then trained to predict the SID of $\boldsymbol{Y}$ by minimizing $\mathcal{L}_{Rec}$
\begin{equation}
\label{eq:Rec}
\mathcal{L}_{Rec}=-\sum_{l=1}^{L}
\log F\left(\boldsymbol{Y}_l \mid \boldsymbol{X}, \boldsymbol{Y}_{<l}\right).
\end{equation}
Here, $\boldsymbol{Y}_l$ denotes the $l$-th token of $\boldsymbol{Y}$, and $F(\cdot)$ represents the conditional probability modeled by the LLM.

\section{Motivation}
\label{sec:motivation}

\begin{figure}[ht]
\centering
\includegraphics[width=0.9\columnwidth]{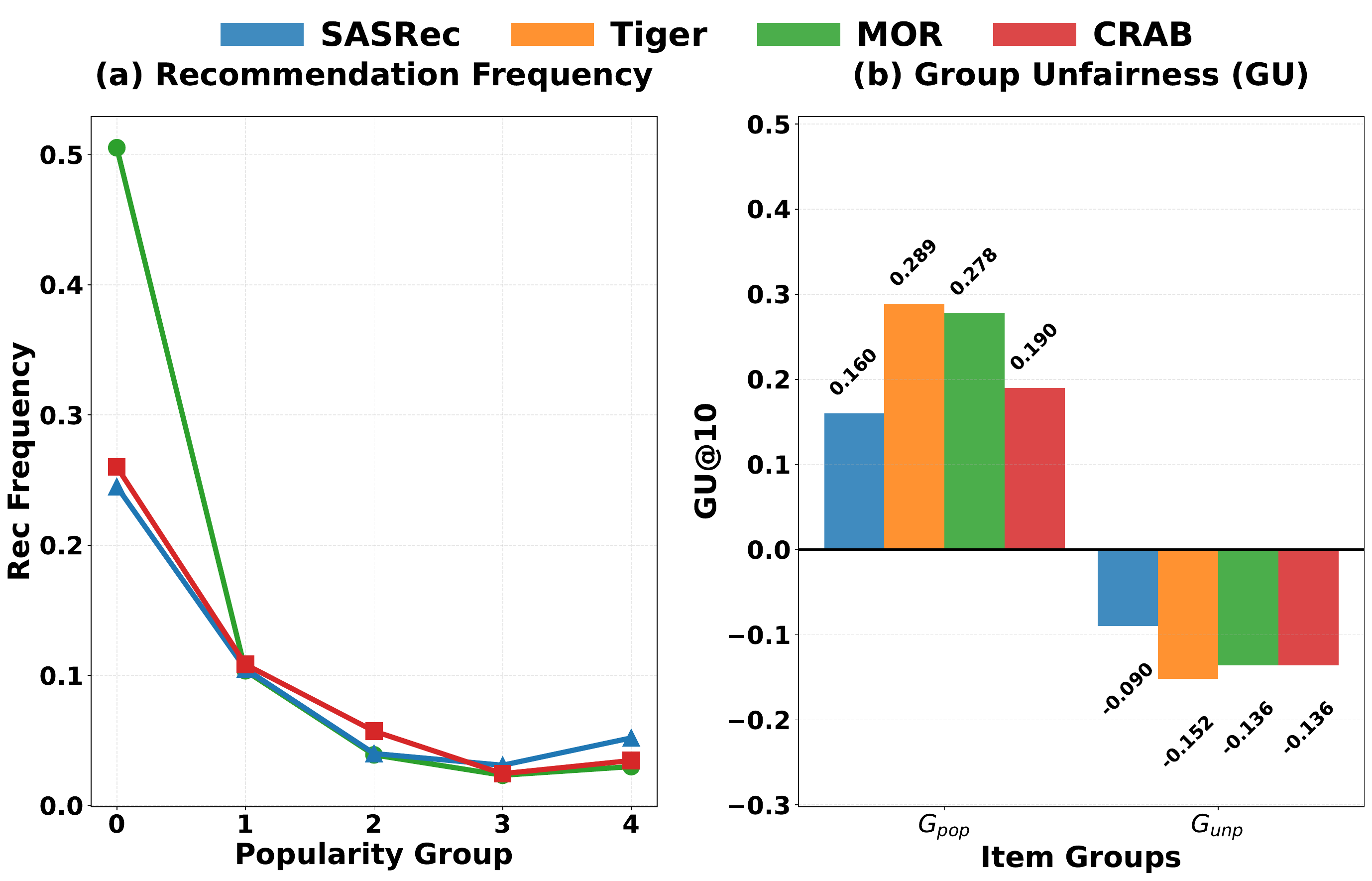}
\caption{Left: Popularity bias of GeneRec on the industrial dataset, with the x-axis representing item groups by popularity.
Right: The \textit{GU} of item groups grouped by token popularity.}
\label{fig:Motivation}
 \vspace{-0.5cm} 
\end{figure}
In this section, we empirically analyze how an imbalanced codebook amplifies item popularity bias in historical interactions. With a slight abuse of notation, let $c_k^l$ denote the $k$-th token in the $l$-th level codebook $\mathcal{C}^l$. We define token popularity as the total frequency of its associated items in the training data. Specifically, given item frequency $f_i$ and the item set $\mathcal{I}_{c_k^l}$ whose $l$-th semantic token is $c_k^l$, the popularity score $P(c_k^l)$ is defined as~\cite{zhu2021popularity,kokkodis2020your}
\begin{equation}
\label{eq:pop}
P(c_k^l)=\sum_{i\in \mathcal{I}_{c_k^l}} f_i,
\quad
\mathcal{I}_{c_k^l} = \{ i \in \mathcal{I} \mid s_i^l = c_k^l \}
\end{equation} 
At each level, we rank tokens by popularity and categorize them into the top $5\%$ over-popular tokens $T_{\text{pop}}$, $5\%-95\%$ neural tokens $T_{\text{neu}}$ and the remaining $5\%$ unpopular tokens $T_{\text{unp}}$.
Items associated with $T_{\text{pop}}$ and $T_{\text{unp}}$ form two groups, denoted as $G_{pop}$ and $G_{unp}$, respectively. We measure \textit{Group Unfairness (GU)}~\cite{jiang2024item}, defined as the discrepancy between recommendation exposure and historical interaction frequency, to quantify bias amplification. As shown in Figure~\ref{fig:Motivation}, tokens in $T_{\text{pop}}$ exhibit significantly stronger bias amplification. On MOR, the GU gap between $G_{pop}$ and $G_{unp}$ reaches $0.42$, indicating that items associated with popular tokens receive disproportionately higher exposure—$1.8\times$ that of SASRec.

Mechanistically, this issue stems from the codebook construction process. Semantically similar items are mapped to the same token; when such items are popular, their interactions accumulate on that token, further increasing its frequency. After training on these token sequences, the recommender becomes biased toward generating items associated with this dominant token, thereby amplifying popularity bias.

\section{Method}
\label{sec:method}
In this section, we introduce the two stages of CRAB respectively.
\begin{figure*}[t]
    \centering
    \includegraphics[width=0.9\linewidth]{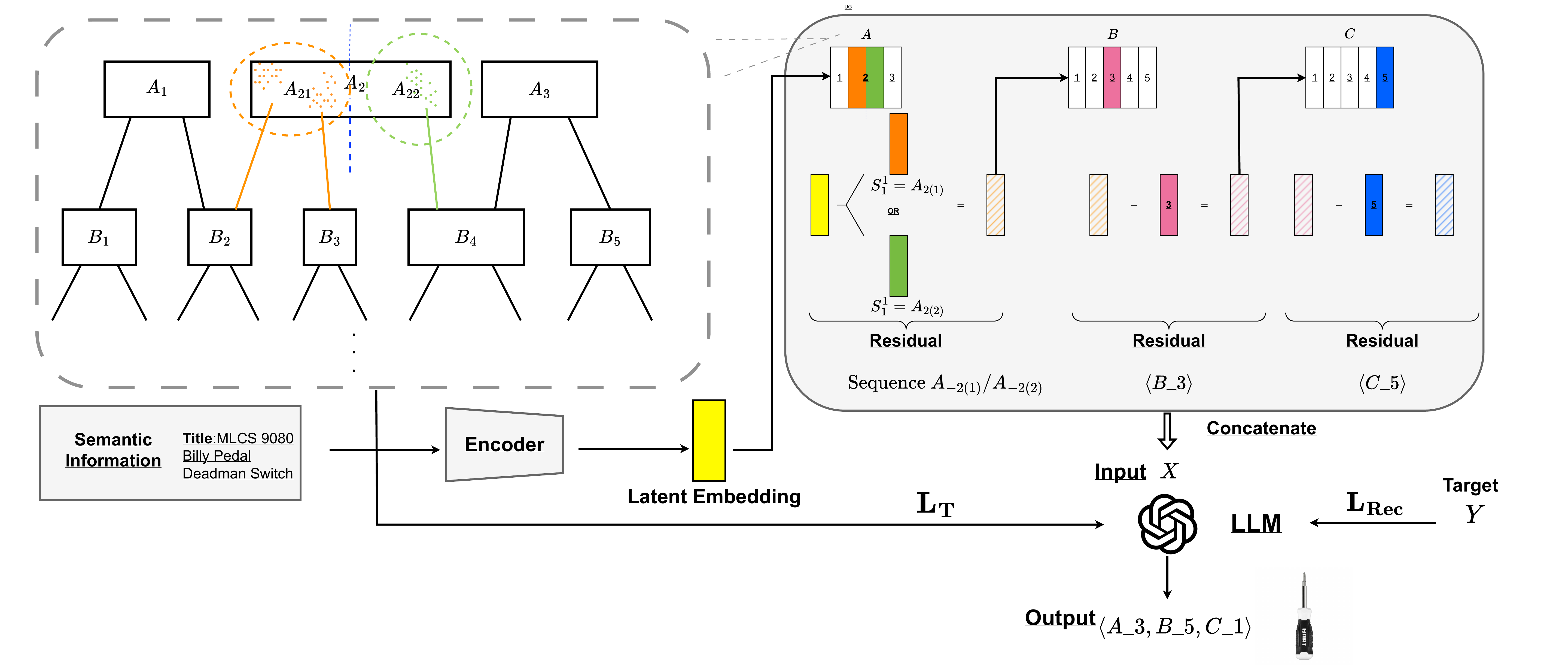}
    \caption{Illustration of CRAB with a three-level codebook in MOR. Over-popular tokens are split by redistributing their child tokens via regularized Kmeans. For clarity, we denote $c_i^{1}$, $c_j^{2}$, and $c_k^{3}$ as $\mathrm{A}_i$, $\mathrm{B}_j$, and $\mathrm{C}_k$, respectively.}
    \vspace{-15pt} 
    \label{fig:CRAB}
\end{figure*}
\subsection{Rebalancing the Codebook}
 The hierarchical token assignment in Equation~\ref{eq:RQ} induces a parent–child relation between tokens at consecutive levels, denoted by $\{c_k^l\rightarrow c_j^{l+1}\}$ if there exists at least one item whose $l$-th and $(l+1)$-th tokens are $c_k^l$ and $c_j^{l+1}$, respectively. Accordingly, for any token $c_k^l \in \mathcal{C}^l$, its children set is defined as
\begin{equation}
\label{lab:child definition}
\mathrm{Ch}(c_k^l)
=
\left\{
c_j^{\,l+1} \in \mathcal{C}^{l+1}
\;\middle|\;
c_k^l \rightarrow c_j^{\,l+1}
\right\}.
\end{equation}

Following Equation~\ref{eq:pop}, we identify over-popular tokens $c_k^l$ and split each of them into $M$ new tokens by clustering the residual representations of its associated items. Since codebook tokens encode hierarchical semantic information from textual descriptions, the splitting procedure should ensure semantic coherence within each newly created token, while preserving the semantic integrity of the $(l\!+\!1)$-th level tokens.
To this end, we further impose a hard constraint that items sharing the same $(l\!+\!1)$-th level semantic token are assigned to the same new token in the $l$-th level. Under the induced tree structure shown in Figure~\ref{fig:CRAB}, this constraint implies that parent tokens are split by redistributing their child tokens, while each child token $c_j^{l+1} \in \mathrm{Ch}(c_k^l)$, together with its associated items, remains intact.

Moreover, to mitigate over-popular tokens, we aim to approximately balance the popularity scores of the newly generated tokens. 
Based on the definition of popularity score(Equation~\ref{eq:pop}), we have  $P(c_k^l)=\sum_{c^{l+1}_j\in\mathrm{Ch}(c_k^l)}P(c^{l+1}_j)$, indicating that the popularity score is preserved between a parent token and its children. 
Formally, for each $c^{l+1}_j\in\mathrm{Ch}(c_k^l)$, we introduce a one-hot vector $\boldsymbol{z}_j \in \mathbb{R}^M$ to indicate the assignment of it to one of the $M$ new tokens. Hence the popularity score for each new token $c_{k(m)}$ can be calculated as follows, and we thereby introducing a balanced loss $\mathcal{L}_{bal}$
\begin{equation}
\label{eq:bal}
P(c^{l}_{k(m)})
=
\sum^{|\mathrm{Ch}(c_k^l)|}\limits_{j=1}\boldsymbol{z}_j[m]P(c^{l+1}_j),\:
\mathcal{L}_{bal}=\sum_{m=1}^{M}\left(P(c^l_{k(m)})-\bar{P}\right)^{2}
\end{equation}
Here $\boldsymbol{z}_j[m] \in {0,1}$ denotes the $m$-th element of $\boldsymbol{z}_j$, and $\bar{P}$ represents the average score. Accordingly, splitting an over-popular token $c_k^l$ can be formulated as a regularized K-means problem that redistributes its child tokens $c_j^{\,l+1} \in \mathrm{Ch}(c_k^l)$ into $M$ new parent tokens.
\begin{equation}
\label{eq:kmeans}
\begin{aligned}
\min_{\boldsymbol{z}} \;\;
& \sum_{m=1}^M
\sum^{|\mathrm{Ch}(c_k^l)|}_{j=1}
\boldsymbol{z}_j[m]n_j
\left\|
\mathbf{\bar{r}}^l_j - \bar{\mathbf{\mu}}_m
\right\|^2+ \lambda\mathcal{L}_{bal},\quad n_j=|\mathcal{I}_{c_j^{l+1}}|\\
\text{s.t.} \;\;
& \boldsymbol{z}_j[m] \in \{0,1\}, \qquad
\sum_{m=1}^M \boldsymbol{z}_j[m] = 1,
\quad
\forall\, c_j^{\,l+1} \in \mathrm{Ch}(c_k^l),
\end{aligned}
\end{equation}
Where $\mathbf{\bar{r}}^l_j$ denotes the mean residual of items $i\in \mathcal{I}_{c_j^{l+1}}$ at the $l$-th level, and $\bar{\mathbf{\mu}}_m$ is the centroid of the newly formed cluster corresponding to $c_{k(m)}$. Eq.~\ref{eq:kmeans} is derived from the variance decomposition in K-means~\cite{blomer2016theoretical}:
\begin{equation}
\label{eq:SSE}
\sum_{i=1}^{n_j} \|\mathbf{r}^{l}_i - \bar{\mathbf{\mu}}_m\|^2
=
\sum_{i=1}^{n_j}
\|\mathbf{r}^{l}_i- \mathbf{\bar{r}}^l_j\|^2
+
n_j \|\mathbf{\bar{r}}^l_j\ - \bar{\mathbf{\mu}}_m\|^2,\: \mathbf{\bar{r}}^l_j=\frac{\sum_{i=1}^{n_j} \mathbf{r}^{l}_i}{n_j}
\end{equation}
Since all $i\in \mathcal{I}_{c_j^{l+1}}$ are assigned to same cluster, the first term in Eq.~\ref{eq:SSE} is constant. Hence, the optimization reduces to the second term.

The regularized K-means objective can be efficiently optimized using the framework of~\cite{raymaekers2022regularized}. Note that Eq.~\ref{eq:bal} holds when the tree structure is strictly maintained in the codebook(e.g., RQ-Kmeans)~\cite{deng2025onerec}. 
To accommodate settings where a child token may have multiple parents (e.g., RQ-VAE)~\cite{kong2025minionerec}, 
we revise Eq.~\ref{eq:bal} by aggregating frequencies only over items associated with both tokens $c^{l+1}_j$ and $c_k^l$:
\begin{equation}
\label{eq:newbal}
P(c^{l}_{k(m)})
=
\sum^{|\mathrm{Ch}(c_k^l)|}\limits_{j=1}\boldsymbol{z}_j[m]P(c^{l+1}_j|c^{l}_k),\:P(c^{l+1}_j|c^{l}_k)=\sum_{i\in \mathcal{I}_{c^{l+1}_j}\cap\mathcal{I}_{c_k^l}}f_i
\end{equation}
\subsection{Hierarchical Semantic Alignment}
Splitting the token $c_k^{l}$ introduces $M$ new tokens. To adapt the LLM to the rebalanced codebook and mitigate bias, we introduce a tree-structure-aware regularizer $\mathcal{L}_{T}$ that promotes representation consistency among tokens sharing the same parent.
\begin{equation}
\begin{aligned}
\mathcal{L}_{T}
=
\sum^{L-1}_{l=1}
\sum^{|\mathcal{C}^{l}|}_{k=1}
\frac{1}{|\mathrm{Ch}(c^{l}_k)|}
\sum_{c \in \mathrm{Ch}(c^{l}_k)}
\left\| e(c) - \bar{e}^l_k \right\|_2^2
\end{aligned}
\end{equation}
where $e(c)$ denotes the LLM embedding of token $c$, and $\bar{e}_k^{l}$ is the mean embedding of the child tokens of $c_k^{l}$. The regularizer applies to both new and existing tokens, serving two purposes: (1) enhancing under-represented tokens via supervision from semantically related siblings, and (2) enabling efficient knowledge transfer to newly introduced tokens after rebalancing.
\begin{table*}[t]
  \centering
  \caption{Performance and Efficiency Comparison on Industrial and Office Datasets.}
  \label{tab:comparison_no_crabe}
  \begin{tabular}{l|cccccc|cccccc}
    \toprule
    & \multicolumn{6}{c|}{\textbf{Industrial}} & \multicolumn{6}{c}{\textbf{Office}} \\
    \textbf{Metric} & \textbf{MOR} & \textbf{Tiger} & \textbf{RW} & \textbf{RR} & \textbf{D$^2$LR} & \textbf{CRAB} & \textbf{MOR} & \textbf{Tiger} & \textbf{RW} & \textbf{RR} & \textbf{D$^2$LR} & \textbf{CRAB} \\
    \midrule
    HR@10 $\uparrow$    & \textbf{0.152} & 0.132 & 0.126 & 0.141 & 0.147 & \textbf{0.152} & \textbf{0.161} & 0.137 & 0.131 & 0.146 & 0.153 & 0.160 \\
    NDCG@10 $\uparrow$  & 0.116          & 0.090 & 0.070 & 0.107 & 0.113 & \textbf{0.117} & \textbf{0.122} & 0.100 & 0.090 & 0.119 & 0.119 & \textbf{0.122} \\
    DGU@10 $\downarrow$ & 0.418          & 0.423 & 0.367 & 0.406 & 0.410 & \textbf{0.356} & 0.423          & 0.427 & 0.386 & 0.410 & 0.414 & \textbf{0.368} \\
    MGU@10 $\downarrow$ & 0.109          & 0.112 & 0.105 & 0.109 & 0.106 & \textbf{0.091} & 0.111          & 0.113 & 0.108 & 0.110 & 0.110 & \textbf{0.093} \\
    \midrule
    Time (h) $\downarrow$ & -           & -     & 3.11  & \textbf{0.21} & 2.75  & 0.28  & -              & -     & 3.25  & \textbf{0.28} & 3.16  & 0.38  \\
    \bottomrule
  \end{tabular}
\end{table*}
\subsection{Model Optimization}
To mitigate popularity bias in GeneRec, we jointly optimize the recommendation loss $\mathcal{L}_{Rec}$ and the hierarchical regularizer $\mathcal{L}_{T}$:
\begin{equation}
\label{eq:all}
\begin{aligned}
\mathcal{L}
=\mathcal{L}_{Rec}+\gamma\mathcal{L}_{T}
\end{aligned}
\end{equation}
where $\gamma$ is a hyperparameter controlling the strength of regularization.
During optimization, we update the embedding layers for both existing and newly introduced tokens.
To improve efficiency, LoRA adapters are applied only to the attention layers of the LLM.
\section{Experiments}

 In this section, we evaluate the performance of CRAB.
\subsection{Experimental Settings} 
\textbf{Datasets}
Experiments are carried out on two real-world datasets Office and Industrial. Following~\cite{kong2025minionerec}, we first filter out users and items with fewer than five interactions. For each dataset, interactions are split chronologically into training, validation, and test sets with an 8:1:1 ratio. 
\textbf{Evaluation Metrics} \label{subsec:metric} Following~\cite{kong2025minionerec}, we evaluate recommendation performance using NDCG@K and HR@K, and measure bias amplification with MGU@K and DGU@K~\cite{jiang2024item}. We also report time cost to assess efficiency.
\textbf{Baselines} \label{subsec:baseline}
To evaluate \textbf{CRAB}, we compare it against two backbone generative models: \textbf{Tiger}~\cite{rajput2023recommender} and \textbf{MiniOneRec (MOR)}~\cite{kong2025minionerec}. We also implement three SOTA popularity debiasing methods atop MOR for a fair comparison: 
(1) \textbf{Reweighting}~\cite{jiang2024item} balances loss contribution via item popularity; 
(2) \textbf{Reranking}~\cite{jiang2024item} penalizes popular items during post-processing; and 
(3) \textbf{D$^2$LR}~\cite{lu2025dual} employs propensity score weighting for LLMs.

\noindent{\textbf{Implementation}}
We implement CRAB based on the MOR framework using Qwen2-0.5B as the backbone. The model is trained on 4$\times$ NVIDIA A100 GPUs for 10 epochs. We employ the AdamW optimizer with a global batch size of 128. The learning rate is set to $1 \times 10^{-4}$ with a weight decay of 0.01. To efficiently train the model, we utilize LoRA~\cite{hu2022lora} with the rank $r=8$, $\alpha=16$. For our codebook rebalancing strategy, we perform a hierarchical split at each level of the codebook with a 10\% splitting ratio. The number of new tokens $M$ is determined by the ratio between the frequency of the target token and the average frequency at the same layer, with an upper bound of $M \leq 3$.

\subsection{Experimental Results}
\textbf{Overall Performance} We compare CRAB with the baselines in Section~\ref{subsec:baseline}, as shown in Table~\ref{tab:comparison_no_crabe}. Overall, CRAB achieves performance comparable to MOR while significantly mitigating popularity bias. Specifically, on the Industrial dataset, CRAB improves HR@10 by \textbf{15.2\%}, \textbf{7.8\%}, and \textbf{3.4\%} over Tiger, RW, and RR, respectively, demonstrating its ability to alleviate bias without sacrificing recommendation quality. In terms of debiasing effectiveness, CRAB achieves the best results, reducing DGU@10 by \textbf{14.8\%} compared to MOR and \textbf{13.2\%} compared to D$^2$LR, while lowering MGU@10 by \textbf{16.5\%} and \textbf{14.2\%}, respectively. Although RW also mitigates bias, it causes severe performance degradation. In contrast, CRAB achieves a better trade-off between accuracy and fairness. Moreover, CRAB is highly efficient, requiring only about \textbf{1/11} and \textbf{1/10} of the training time of RW and D$^2$LR. Although RR is more efficient due to re-ranking, CRAB achieves superior performance and fairness.

\begin{figure}[ht]
\centering
\includegraphics[width=0.83\columnwidth]{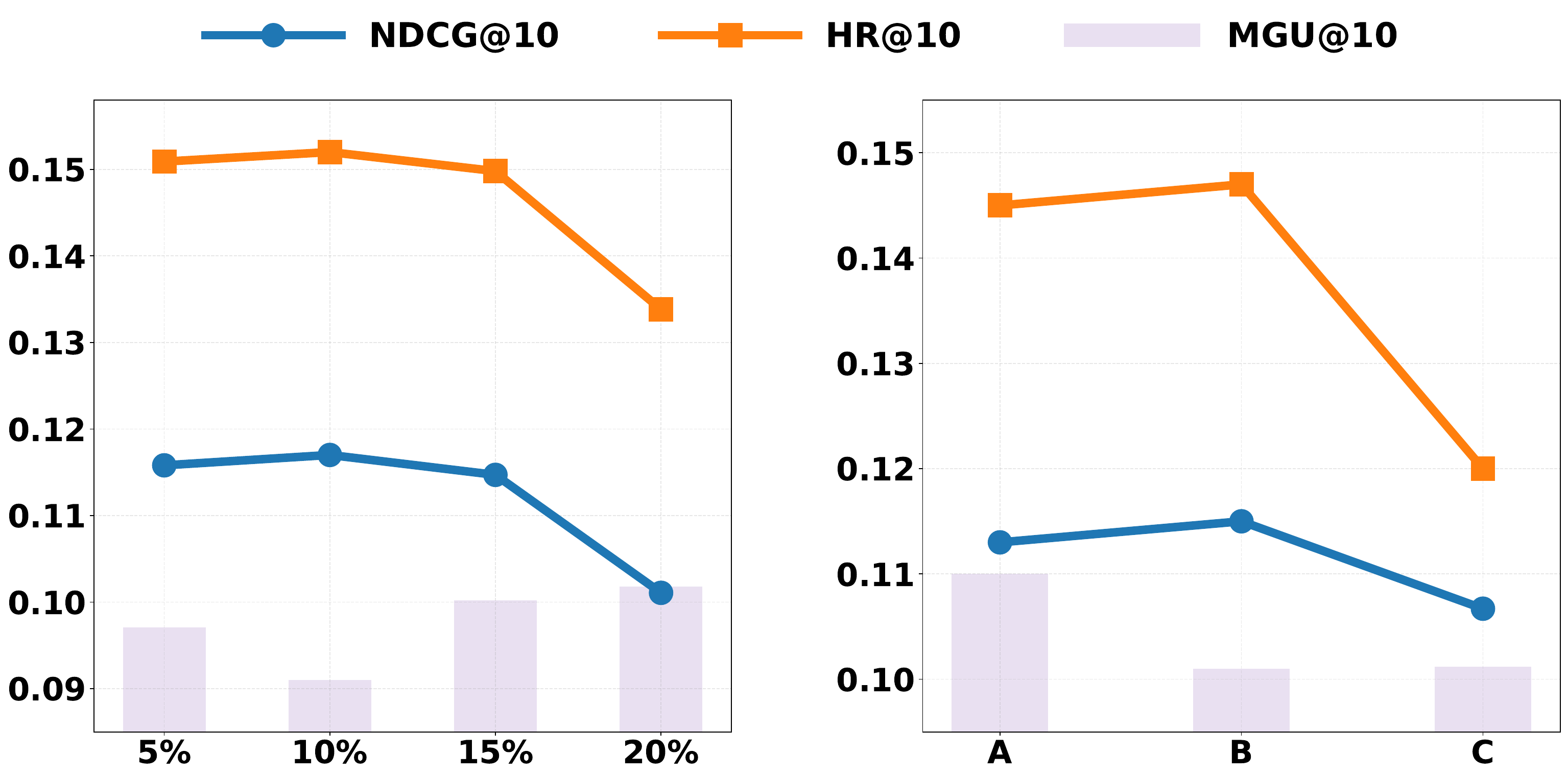}
\caption{Left: MOR performance under different splitting ratios ($5\%$–$20\%$) at each level on the Industrial dataset. Right: MOR performance when splitting the top $5\%$ most popular tokens at levels A, B, and C separately.}
\label{fig:Deep}
 \vspace{-0.5cm} 
\end{figure}

\begin{figure}[ht]
\centering
\includegraphics[width=0.88\columnwidth]{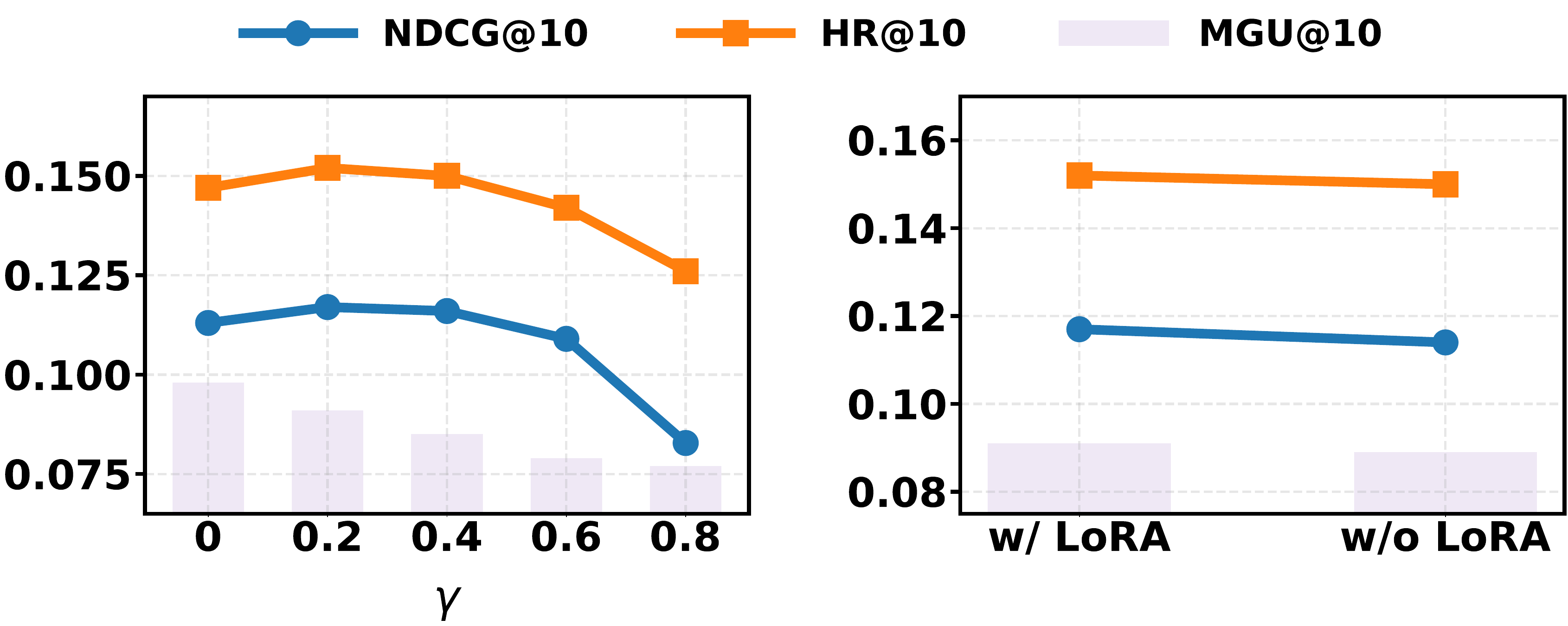}
\caption{Left: Effect of $\gamma$ Right: Effect of LoRA}
\label{fig:Ablation}
 \vspace{-0.5cm} 
\end{figure}

\subsection{In-depth Analysis}
\noindent{\bf Impact of Splitting Ratio} We analyze how the proportion of tokens to split at each level affects CRAB. As shown in the left part of Figure~\ref{fig:Deep}, as the proportion increases, the performance first improves and then declines, showing a similar trend to bias amplification. This suggests that splitting a small fraction(less than $10\%$) of popular tokens can smooth token popularity and enhance the representation of unpopular tokens, thereby improving long-tail items exposure. However, over-splitting tokens may disrupt the semantic integrity of the codebook and lead to performance degradation.

\noindent{\bf Impact of Splitting Position} We analyze splitting at different levels by splitting only the top-$5\%$ popular tokens from level A, B, and C in MOR separately. From the right part of Figure~\ref{fig:Deep}, splitting over-popular tokens at level B improves representation while significantly mitigating bias, suggesting that intermediate-level tokens concentrate excessive semantic information, consistent with the “Hourglass” phenomenon~\cite{kuai2024breaking}. In contrast, splitting only the last level may hurt performance, as its semantics are already fine-grained.

\noindent{\bf Ablation Study}
We first examine the effect of the hyperparameter $\gamma$, which balances the recommendation objective and representation consistency. As shown in Figure~\ref{fig:Ablation}(left), as $\gamma$ increases, CRAB places more emphasis on $\mathcal{L}_{T}$. When $\gamma \leq 0.2$, the recommendation performance remains stable. However, for $\gamma > 0.2$, NDCG drops sharply. To achieve a proper trade-off, we set $\gamma = 0.2$. We then evaluate the necessity of LoRA and observe that removing LoRA leads to noticeable drops in both NDCG and HR from the Figure.

\section{Conclusion}
In this paper, we present the first systematic investigation of popularity bias in GeneRec and propose CRAB, a novel method that mitigates bias by rebalancing the codebook.
\bibliographystyle{ACM-Reference-Format}










\end{document}